\documentclass{aa}
\usepackage[dvips]{graphicx,color}

\begin{document}

%
   \title{UV observations of the galaxy cluster Abell 1795 with the optical
   monitor on XMM-Newton\thanks{Based on observations
   obtained with XMM-Newton, an ESA science mission with instruments
   and contributions directly funded by ESA Member States and the USA
   (NASA).}}

   \subtitle{}

   \authorrunning{J.P.D.Mittaz et al.}
   \titlerunning{UV observations of the galaxy cluster Abell 1795 with the
   optical monitor}

   \author{J.P.D.Mittaz\inst{1}, 
   J.S.Kaastra\inst{2}, T.Tamura\inst{2},
   A.C.Fabian\inst{3}, R.F.Mushotzky\inst{4},
   J.R.Peterson\inst{5},  
   Y.Ikebe\inst{6}, D.H.Lumb\inst{7}, F.Paerels\inst{5},
   G.Stewart\inst{8} 
   \and S. Trudolyubov\inst{9} 
	}
	
   \offprints{J. Mittaz}
   \institute{Department of Space and Climate Physics, University College London,
	Mullard Space Science Laboratory, Holmbury St. Mary, Surrey, U.K.
        \and
	SRON Laboratory for Space Research, Sorbonnelaan 2, 3584 CA, Utrecht,
	The Netherlands
	\and
	Institute of Astronomy, University of Cambridge, Madingly Road,
	Cambridge, CB3 0HA, U.K.
	\and
	NASA/GSFC, Code 662, Greenbelt, MD20771, U.S.A.
	\and
	Columbia Astrophysics Laboratory and Department of Physics, Columbia
	University, 538 West 120th Street, New York, NY 10027, U.S.A.
	\and
	Max-Planck-Institut f\"{u}r extraterrestrische Physik,
	Postfach 1312, 85741 Garching, Germany
	\and
	Atrophysics Division, European Space Agency, ESTEC, Postbus 299,
	2200AG Noordwijk, The Netherlands	
	\and
	Department of Physics and Astronomy, The University of Leicester,
	Leicester, LE1 7RH, U.K.
	\and
	Los Alamos National Laboratory, NIS-2, Los Alamos, NM 87545, U.S.A.
             }
   \date{Received ; }

\abstract{ 
We present the results of an analysis of broad band UV observations of the
central regions of Abell 1795 observed with the optical monitor on XMM-Newton.
As have been found with other UV observations of the central regions of
clusters of galaxies, we find evidence for star formation.  However, we also
find evidence for absorption in the cD galaxy on a more extended scale than has
been seen with optical imaging.  We also report the first UV observation of
part of the filamentary structure seen in H$\alpha$, X-rays and very deep U
band imaging.  The part of the filament we see is very blue with UV colours
consistent with a very early (O/B) stellar population.  This is the first
direct evidence of a dominant population of early type stars at the centre of
Abell 1795 and implies very recent star formation at the centre of this
cluster.
}

\maketitle

%

\section{Introduction}

The final fate of material in the centre of cooling flow clusters remains a
mystery.  Perhaps the most obvious end point for matter cooling out of the
cooling flow is in stars, and it is indeed true that the cD galaxies
in cooling flow clusters often have anomalously blue optical colours
(e.g. Allen 1995), implying higher rates of star formation than normal.  There
is also a reported correlation between the amplitude and radial extent of the
colour anomalies and the inferred mass accretion rate from X-ray measurements
(McNamara \& O'Connell 1992).  However, much of the observations of star
formation in clusters of galaxies is based on optical measurements, and so
crucial information about the emission from early type, and therefore young,
stars is missing.  Unfortunately, UV observations of the cores of cooling flow
clusters, where the emission from early type stars would be strongest, are
relatively sparse (e.g. A1795 Smith et al. 1997, CL 0939+4713 Buson et
al. 1998).

The cD galaxy in Abell 1795 has been extensively studied and shows a number of
remarkable features, including extended nuclear emission line gas (Cowie et al
1983; Hu, Cowie \& Wang 1985; Heckman et al. 1989), an unusual blue continuum
(Hu 1992; Allen 1995) and filamentary structure in X-rays (Fabian et al. 2000).
These properties have made it one of the best candidates for star formation
arising from a cooling flow.  Previous UV studies with UIT (Smith et al. 1997)
have shown the presence of emission from early type stars in the cD galaxy with
data that were consistent with either continuous star formation or a recent (4
Myr) burst of star formation.  

Here we report on new multi-filter UV observations of the central 8 arcminutes
of the cooling flow cluster Abell 1795 taken as part of an XMM-Newton PV
observation.  Other aspects of the XMM-Newton observations will be discussed
elsewhere (Tamura et al. 2001, Arnaud et al. 2001)

\section{OM data reduction}

Abell 1795 was observed in the U (3000\AA - 4000\AA), UVW1 (2400\AA - 3600\AA),
UVM2 (2000\AA - 2600\AA) and UVW2 (1800\AA - 2400\AA) filters (for more details
of the optical monitor see Mason et al. 2001).  The observations used the
default configuration where the whole field of view is sampled using five
individual exposures.  Most of the field of view was sampled with 1 arcsecond
pixels, but a small section (110 arcseconds by 110 arcseconds) of the central
region of the field was continuously monitored at higher resolution (0.5
arcsecond pixels).  This means different regions of the field of view have
different total exposure times.  The outer regions of the
detector have the lower exposure times of U: 1.5 ksec, UVW1: 2.5 ksec, UVM2:
3.0 ksec, UVW2: 3.98 ksec.  The central portions of the detector have longer
exposure times of U: 7.5 ksec, UVW1: 12.5 ksec, UVM2: 15.0 ksec, UVW1: 19.9
ksec.

The data were processed using the SAS (v4.1) tasks written to analyse optical
monitor data.  To maximise the signal-to-noise we have summed all the data in a
given detector window (area on the sky).  This can be important in the UV where
an individual exposure does not have enough counts to enable removal of mod-8
noise (see Mason et al. 2001).  The sources were then detected on each
individual window (using the SAS task OMDETECT).  and the output sources lists
where then edited by hand to remove sources caused by stray light in the OM (for
details of instrumental artifacts see Mason et al. 2001).  Further, sources
whose background was strongly affected by stray light were re-extracted using an
annular background estimate as a more robust method.  This mainly affects the
central few arcminutes of the field.  The observed count rates were then
converted into magnitudes using the derived OM calibration.  In the case of the
U filter, the instrumental magnitude was converted into Johnson U using a
colour correction.  Since there is no corresponding standard for the far UV
filters, UVW1, UVM2 and UVW2 were kept in instrumental magnitudes.  The
estimated errors on the magnitude conversion are less than 0.05, derived from
the latest OM calibration files.

In total, 284 sources were detected in the full $17 \times 17$
arcminute field of view and the count rates were then converted to instrumental
magnitudes.  The identification of the sources via correlation with catalogs
and detailed galaxy data will be presented in a follow-up paper.

\section{Results}

\subsection{Colour-colour plots}

\begin{figure*}
\begin{center}
\includegraphics[width=14cm]{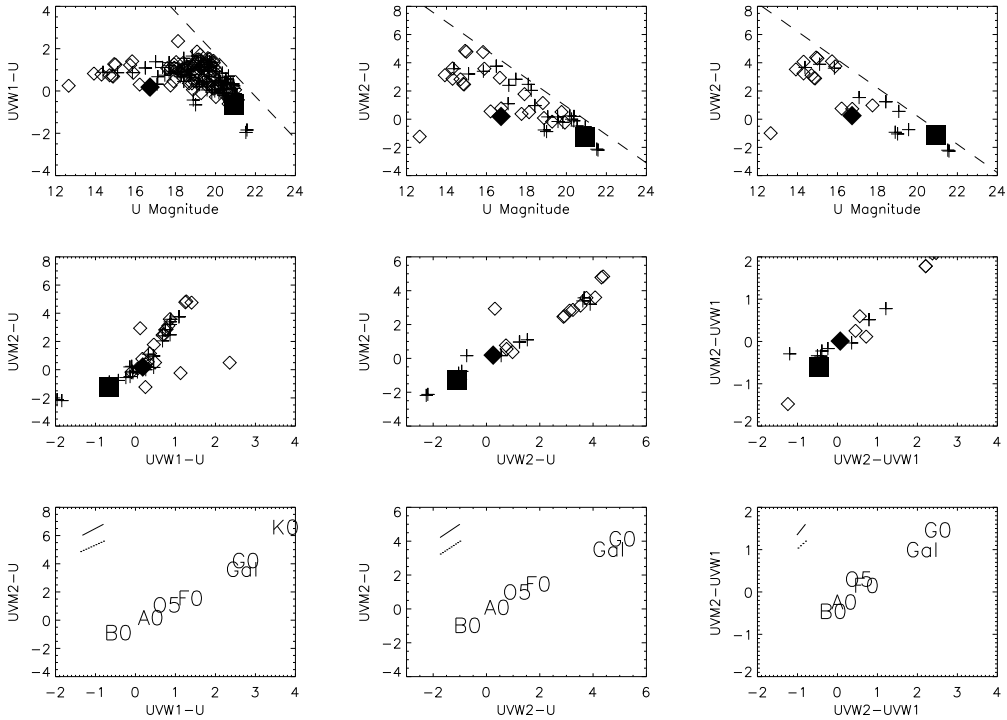}
\end{center}
\caption{The magnitude-colour plots and colour-colour plots for the Abell 1795
field.  The top three panels show magnitude-colour plots together with the
detection limits (dashed lines).  The middle three panels show the
colour-colour plots where point-like sources are represented by crosses,
extended sources are represented by open diamonds, and the cD galaxy and UV
blob are showed as a filled diamond and square respectively.  The lower three
panels show the position of different stellar types in the colour-colour plane.
The solid lines show the effect of Galactic extinction and dotted lines show
the effect of a dusty galaxy extinction, both up to an E$_{B-V}$ of 0.22 (Allen
1995).}
\label{color-color}
\end{figure*}

With our final sample of 284 sources we have constructed magnitude-colour and
colour-colour plots (see Figure~\ref{color-color}).  In the plots we have
discriminated between point like and extended sources by looking at the
distribution of the size of the major axis of the derived source ellipse.
Since the measured FWHM in the UV is approximately 3 arcseconds anything with a
FWHM greater than 4 arcseconds is considered extended.  Point like sources are
shown as crosses and extended sources are shown with a diamond. Two objects of
special interest are shown by other symbols.  The cD galaxy is represented by
the filled diamond, and the emission feature 20 arcseconds south of the cD
galaxy is represented by a square (see section~\ref{sec-cD} and
\ref{sec-blob}).  The derived detection limits determined from the background
in each observation for each filter were U = 21.67, UVW1 = 21.74, UVM2 = 20.89
and UVW2 = 20.22.  On the magnitude-colour plots the derived detection limits
are shown as dashed lines and there is good agreement between the observed
detection limits of the sources and the derived detection limits.

The colour-colour plots show the range of UV colours exhibited by the sources.
The majority of the sources lie on well defined lines and the lower three
panels show the position in the colour-colour plots of different stellar types
(taken from Pickles (1998)) as well as a typical field galaxy spectrum.  For
clarity, the far UV colours for different stellar types are also listed in
table~\ref{tablecol}.  The apparently anomalous position of the O5 star is
because the UV filters bracket the strong absorption dip at 2200\AA caused by
the presence of a wind.  All the spectra have been redshifted to the cluster
redshift but do not include any correction for extinction.  Marked on the
plots are two curves showing the effect of reddening by a Galactic (solid) and
a dusty galaxy (dotted) line (Witt et al. 1992).  Most of the objects lie on or
near the loci defined by the change in stellar spectral type.  In general the
point sources have the colours of F and G stars, and while it is unlikely that
the extended sources will have single stellar like spectra, it at least allows
an estimation of the dominant stellar type in the galaxies.  As might be
expected with the galaxies at the centre of Abell 1795, many of the extended
sources have `red' UV colours consistent with red galaxies.  There are,
however, a number of very blue points including a number of extended sources.
The extended source seen at the extreme right in the UVW1-UVW2 and UVW1-UVM2
plot is a very bright galaxy affected by coincidence losses (see Mason et
al. 2001).  The cD galaxy and its associated UV blob also lie to the extreme
right of the colour-colour plots.

\begin{table}
\begin{center}
\begin{tabular}{|l|c|c|} \hline
 & UVW2 - UVW1 & UVM2 - UVW1 \\ \hline
O5 & 0.192 & 0.133 \\
B0 & $-0.454$ & $-0.628$ \\
A0 & $-0.141$ & $-0.409$ \\ 
F0 & $-0.405$ & $-0.018$ \\
G0 & 2.170 & 1.294 \\
K0 & 3.592 & 2.674 \\ \hline
\end{tabular}
\end{center}
\caption{The OM far UV colours of different stellar types}\label{tablecol}
\end{table}

\subsection{The cD galaxy}\label{sec-cD}

\begin{figure*}
\begin{center}
\includegraphics[width=12.5cm]{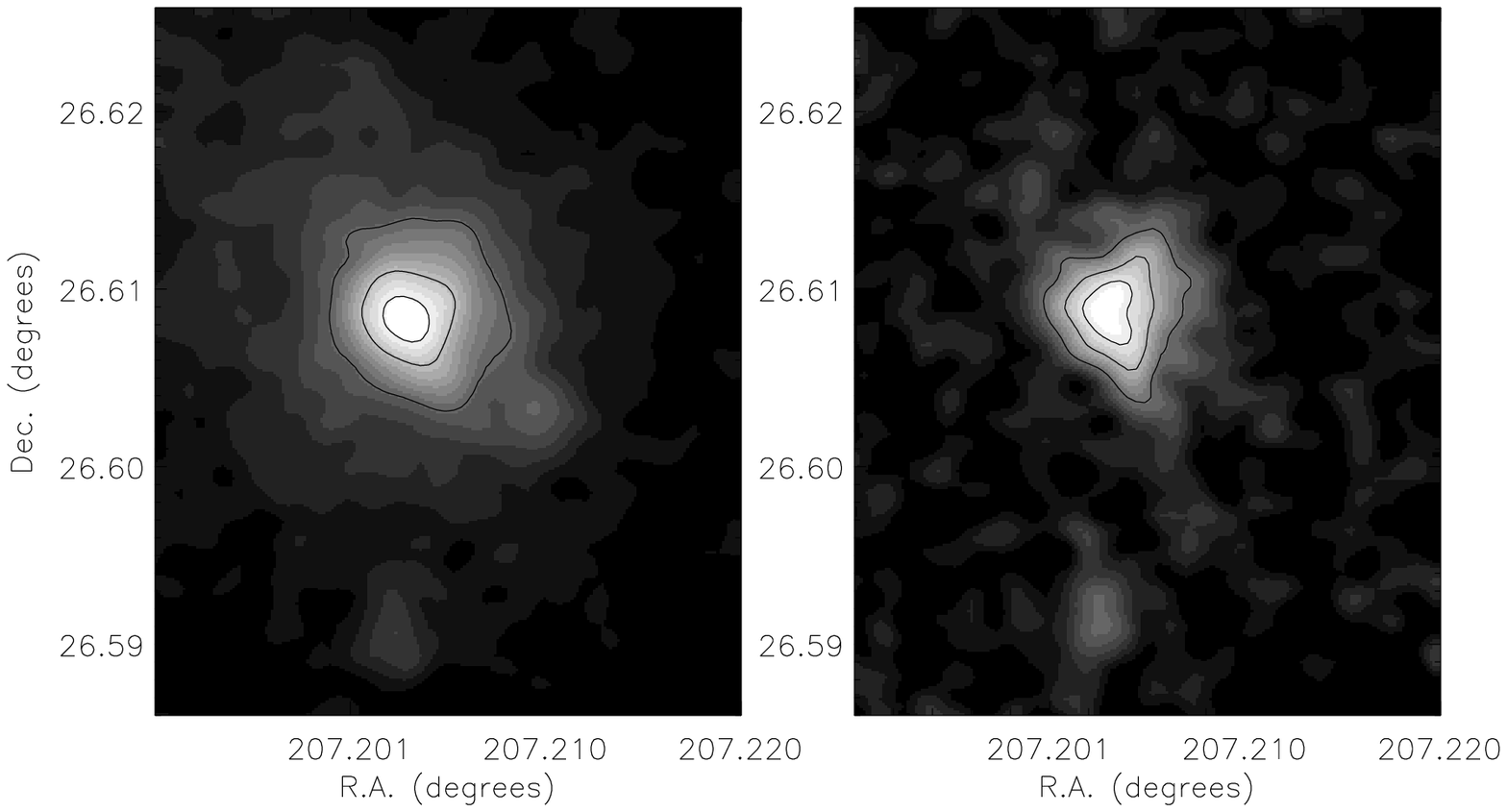}
\end{center}
\caption{The U and UVW2 images of the central regions of Abell 1795. The size
of each image is 35 by 48 arcseconds with the cD galaxy located near the middle
of each image and the UV blob approximately 20 arcseconds to the south of the
cD.  The cD galaxy in the two filters appears to show different morphologies.}
\label{cD}
\end{figure*}

Because we have colour data on the UV emission from the cD we can set limits on
the number of young stars.  If we make the assumption that all or most of the
UVW2 emission (the bluest filter) arises from O stars (as was done by Smith et
al. (1997)), then the estimated number of stars required to give the observed
flux is approximately 12000.  This estimate is unlikely to be affected by
emission from an old stellar population since the effect is small, less that
2\%.  This is in approximate agreement with 18000 O stars estimated by Smith et
al. (1997), where their data was taken in a far UV filter centred at 1520\AA\
much further in the UV than UVW2, our bluest filter.  However, both estimates
are likely to be upper limits, since from figure~\ref{color-color} we would
estimate that the cD galaxy has a UV spectrum redder than that of an early B or
late type O star.

Figure~\ref{cD} shows an image in sky coordinates of the cD galaxy taken in the
U and UVW2 filters.  In the U band the source shape is consistent with the data
of McNamara \& O'Connell (1993), with an ellipticity of 0.6 and position angle
of 17 degrees.  However, in detail there are likely to be problems with the
scattered light in OM which make the background subtraction a non trivial
exercise.  Fortunately, the effect of this scattered light is strongly filter
dependent and there are no serious background subtraction problems with the far
UV filters.  Looking in the far UV, the UVW2 image shows a relatively sharp
discontinuity running north-source to the right of the cD galaxy centre.  The
only other UV image published on the central galaxy (Smith et al. 1997) also
show a similar feature, although the spatial resolution of the data taken with
the UIT was of the order of $6.8''$, about twice the nominal FWHM of OM in the UV
filters ($\sim 3''$).  Smith et al. explained this feature as due to the dust
absorption seen in the HST data (Pinkney et al. 1996) but figure~\ref{cD} shows
that it appears on a larger scale than the dust lanes seen in the HST data.

\begin{figure}
\includegraphics[width=8cm]{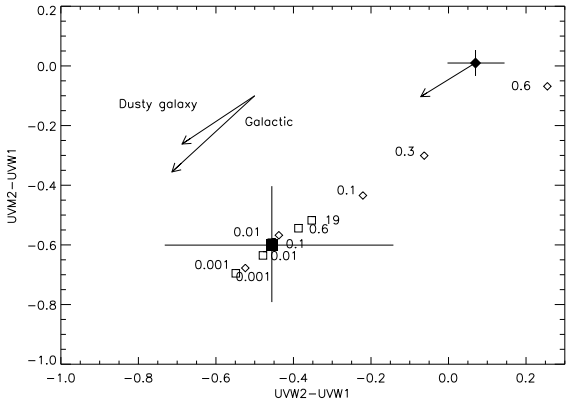}
\caption{Colour-colour plot showing the position of the UV blob (filled square)
and the cD galaxy (filled diamond) together with the track of a single burst of
star formation (open diamonds), or constant star formation rates (open squares)
evolving with time.  The arrows show the effect of extinction (both
Galactic and a dusty galaxy) up to an E$_{B-V}$ of 0.22 (Allen 1995).
The arrow associated with the cD galaxy shows the effect of removing an old
stellar population spectrum.}
\label{starform}
\end{figure}

The effect of dust affects our conclusions regarding the star formation history
of the cD galaxy.  Hu (1992) estimated the intrinsic extinction to the filament
as $E_{B-V} = 0.14$ but with a wide range of possible values ($0.02 < E_{B-V} <
0.22$).  Allen (1995) has suggested that the value is closer to 0.22.  Smith et
al. (1997) concluded that after taking into account the effect of dust their UV
data was consistent with either an old stellar population together with a
single burst of star formation occurring approximately 2-5 Myr ago, or constant
star formation over the last 5-10 Gyr.  Figure~\ref{starform} shows the
position of the cD galaxy in the UVW2-UVW1, UVM2-UVW1 plane together with the
position of different star formation models taken from Bruzual \& Charlot
(1993).  We have concentrated on the far UV data to avoid any problem with the
scattered light.  For both models the different points on the plot
corresponding to different ages after the burst of star formation with the
relevant time shown next to the symbol.  Also plotted on figure~\ref{starform}
are the lines of extinction from two different extinction laws, Galactic and a
dusty galaxy model (Witt et al. 1992), and the arrow associated with the cD
galaxy shows the effect of removing an old stellar population on the position
of the cD galaxy. The end point of this track is where all the U band emission
is from the old stellar population, an overestimate since the cD galaxy is
known to be bluer than a standard elliptical galaxy.  Unlike Smith et
al. (1997), our far UV colours only seem to be consistent with a burst of star
formation sometime between 50 and 200 Myrs ago (depending on extinction and
removal of an old stellar population), and not with continuous star formation.
However, all our UV filters are not as blue as the UV filter used by Smith et
al., so it is possible that the UV spectrum is more complex than the simple
models used here.


\subsection{The UV blob}\label{sec-blob}

On the colour-colour plots (Figure \ref{color-color}) the bluest extended
source in the whole field lies just south of the cD galaxy.  This object can be
seen in figure~\ref{cD} as a blob of very blue emission approximately 20
arcseconds south of the cD galaxy.  Its approximate coordinates are R.A.: 13 48
52.8 DEC: +26 35 33 which is accurate to 2-3 arcseconds.  This blob is almost
certainly the same feature as blob E seen in McNamara et al (1996) and it is
therefore likely that it is part of the same filamentary structures seen in the
deep U band imaging of McNamara et al (1996).  However, with the OM data we
can, for the first time, get estimates of its UV colours.  Unlike the case for
the cD galaxy, the colours of the blob (the filled square) are much more
consistent with an early star spectrum (early B or late O star), implying that
star formation is more recent in this object.  Figure~\ref{starform} shows the
far UV colours of the UV blob compared with different star formation models.
Given the relatively large error bars the UV colours are consistent with both
types of extinction law together with a burst of star formation that started
before 100 Myr ago, or continuous star formation.  If we use the extinction
assumed by Smith et al (1997) (E$_{B-V}$ = 0.14), then the maximum time scale for
a single burst of star formation drops to within 10 Myrs, or to 100 Myr for
continuous star formation.  Since the blob is very blue in all colours, the
implication is that star formation is very recent, with little evidence for an
evolved stellar population, and must be related to some recent occurrence or
phenomena at the centre of Abell 1795.

Again we can estimate the number of O or B stars required to give the observed
flux, which is estimated as 1130 O stars or 4980 B stars.  The implied UV
luminosity from the O stars is then approximately $1.2 \times 10^{43}$ ergs
s$^{-1}$.  Of course again this is an upper limit to the UV luminosity since it
assumes only O stars contribute to the flux.  From these numbers we can then
estimate a star formation rate of $\sim 10 M_\odot$ yr$^{-1}$ (assuming the IMF
of Miller \& Scalo (1979)).  This is still much less than the mass deposition
rate from analysis of previous X-ray data of $\sim 130 M_\odot$ yr$^{-1}$
(Fabian et al.  1994, Tamura et al. 2001).  Again, extinction is very important
in the UV and based on the extinction estimates (see section~\ref{sec-cD}) we
can increase the observed luminosities and therefore inferred star formation
rates by a factor of 3, but still less than the X-ray rate.

There is a well known H$\alpha$ filament to the south of the cD galaxy
(e.g. Cowie, Hu, Jenkins and York 1983) which is aligned in a north south
direction.  This blob lies close to the peak of the H$\alpha$ emission along
the filament and so may be associated with it.  However, comparing the relative
fluxes makes it unlikely that these stars are solely responsible for the
H$\alpha$ emission.  At the peak of the H$\alpha$ emission the estimated flux
was $\sim 2 \times 10^{-16}$ ergs cm$^{-2}$ s$^{-1}$ arcsec$^{-2}$ for the
H$\alpha$ + [N II] $\lambda 6568$\AA\ complex.  The approximate total size of
the UV blob is 20 arcsec$^2$ implying that the total H$\alpha$ + [NII]
emission associated with the blob is $\sim 4 \times 10^{-15}$ ergs cm$^{-2}$
s$^{-1}$ or a luminosity of $7 \times 10^{40}$ ergs s$^{-1}$.  Using the numbers
quoted in Allen (1995) for the amount of H$\alpha$ emission from a single O5
star ($\sim 6 \times 10^{36}$ ergs cm$^{-2}$ s$^{-1}$) gives $5 \times 10^{39}$
ergs s$^{-1}$ for all the stars in the blob.  If we include the effect of
extinction we would increase the observed UV flux and therefore the estimated
luminosity of H$\alpha$, which then becomes $1.5 \times 10^{40}$ ergs
cm$^{-2}$ s$^{-1}$.  This is still lower than the observed flux by a factor of
four or so.  Coupled with the fact that the estimation of the H$\alpha$
emission can be considered to be only an upper limit, it seems unlikely that
all the H$\alpha$ emission can be powered by the observed UV flux seen in the
OM data.

\section{Discussion}

\subsection{The cD galaxy}


As has been reported by Smith et al. (1997), UV data on the cD galaxy indicate
that there is a population of young stars.  The OM data is consistent with
their result of a single burst of star formation, although the estimated
timescale for the occurrence of this burst is longer than that derived by Smith
at al. (1997).  However, the OM filters and the one used by Smith et al. cover
different wavebands, so any difference may be due to the object having a more
complex spectrum than the simple models used for comparison.  The cD galaxy
also appears to have quite a lot of structure with an apparent colour gradient
across the galaxy together with a discontinuity running north south seen in the
far UV filters.  A possible explanation for the discontinuity is obscuration by
dust or gas blocking out the UV emission in that area.  An estimate of the
E$_{B-V}$ required to remove the UVW2 emission is $\sim 0.16$, twice the
measured E$_{B-V}$ for the central portions of Abell 1795 as a whole.  We
therefore may be seeing obscuring material on much large scales than the
reported dust lanes seen with HST data (Pinkney et al. 1996).  While the
resolution of OM is insufficient to make a direct link between the HST data and
OM data, it is possible that the UV obscuration seen with OM is related to
these dust lanes, since both the dust lanes and the UV absorption lie on the
same side of the cD galaxy.  The OM data may therefore be tracking dust or gas
out to a larger radius than was possible with the HST data because the UV
filters are uniquely sensitive to extinction.

\subsection{The UV blob}

The UV blob is potentially the most interesting object seen in the data.  For a
start it gives us the possibility of studying the star formation at the centre
of a cooling flow cluster without the central cD galaxy getting in the way.
The UV colours imply not only the presence of young stars, but there is little
evidence of an evolved stellar population expected from an
elliptical galaxy.  It is therefore possible that the blob is the consequence
of recent star formation alone, without a pre-existing stellar population.  It
is also located very close to, or within the filamentary structure seen in both
H$\alpha$ and deep U band imaging, although since the positions obtainable
from OM at the present moment are somewhat uncertain an exact positional match
is not possible.  From the measurements discussed above it seem unlikely that
all of the H$\alpha$ emission is powered from the UV light seen in the OM
data, with at least a factor of 4-5 discrepancy between the observed flux and
that required for the H$\alpha$ emission.  However, some of the H$\alpha$
emission may arise from shocked gas (e.g. Anton 1993) and it has been suggested
that cooling material from the cooling flow could contribute to the ionising
flux (e.g. Voit \& Donahue 1990).  It is also not clear from the OM data
whether there is further UV emission along the tail since there is not
sufficient signal-to-noise.  The emission at the peak of the blob in the UVW2
filter is at about $1.8 \times 10^{-3}$ counts s$^{-1}$ arcsec$^{-1}$ or 7
$\sigma$ above background so a factor of 2 decrease in flux would make it hard
to see any residual emission.  In the other filters, such as UVW1, the
significances are greater but the scattered light component is also stronger
making it impossible to be sure than any connecting emission is real.

There is also X-ray emission associated with the H$\alpha$ filament which has
been seen by Chandra (Fabian et al. 2000).  Not only is there filamentary
structure seen in X-rays but it also appears that the brightest point on the
X-ray filament is coincident with the centroid of the overall X-ray emission.
Comparison with the Chandra data then implies that the UV blob is also
coincident with this point and so the UV blob is at or near the centre of the
gravitational potential.  At this location we might expect a general reservoir
of matter which can form stars.  On the other hand the cD galaxy appears offset
relative to the centre gravitational well and observations have also shown that
the cD galaxy does not appear to be at rest in the gravitational potential of
the cluster but has a peculiar velocity of about 150 km s$^{-1}$ (Oegerle \&
Hill 1994).  Therefore, it is likely that star formation has occurred very
recently in this source (see figure~\ref{starform}) and that this star
formation could have been triggered by the motion of the cD galaxy.  If we
assume as Smith et al. did that the extinction is close to an E$_{B-V}$ of
0.14, the data implies that a single burst would have occurred before 10
million years ago.  The approximate minimum distance of the cD galaxy from the
UV blob is approximately 24 kpc setting a lower limit to the estimated average
velocity of 2300 km s$^{-1}$ or for continuous star formation models of 230 km
s$^{-1}$.  The observed radial velocity is $\sim 150$ km s$^{-1}$.  It is
therefore possible that the motion of cD galaxy could have shocked gas,
triggering either star formation in a single burst or continuously.
Of course, in order for stars to form there must be sufficient cold
material to generate stars, perhaps generated by a cooling wake (e.g. David et
al. 1994) or from the cooling flow itself.


\section{Conclusions}

The XMM-Newton OM data have illustrated the usefulness of UV observations in
studying populations of early type stars in clusters of galaxies.  We have set
constraints on models of star formation at the centre of Abell 1795 and have
also detected the presence of extra extinction in the cD galaxy.  We have also
presented the first UV observations of emission from the filament
seen in deep U band images and in H$\alpha$ to the south of the cD galaxy.  In
particular the observations of the central regions of Abell 1795 have
highlighted a population of early type stars located at or near the centre of
the gravitational potential that is most likely the consequence of recent star
formation.  While the exact significance of this in terms of detailed models of
material cooling out of the intercluster medium is not yet clear, these data
should, in combination with optical/IR data, provide constraints on the
timescale and rate of star formation from the intercluster medium itself.



\end{document}